# Fabrication of InP nano pillars by ECR Ar ion irradiation

D. Paramanik[1,2], T. T. Suzuki[1], N. Ikeda[1], Y. Sugimoto[1], T. Nagai[1], M. Takeguchi[1] and C Van Haesendonck[2]

1) National Institute for Materials Science, 1-2-1 Sengen, Tsukuba, Ibaraki 305-0047, Japan
2) Laboratory of Solid state Physics, Katholiek University of Leuven, Celestijnenlaan 200 D - Box 2414, 3001 Heverlee (Leuven), Belgium

## Abstract

Regular arrays of InP nano pillars have been fabricated by low energy Electron Cyclotron Resonance (ECR) $Ar^+$ ion irradiation on InP(111) surface. Several scanning electron microscopy (SEM) images have been utilized to invetsigate the width, height, and orientation of these nano pillars on InP(111) surfaces. The average width and length of these nano-pillars are about 50 nm and 500 nm, respectively. The standing angle with respect to the surface of the nano-pillars depend on the incidence angle of the Ar ion irradiation during the fabrication process. Interestingly, the growth direction of the nano pillars are along the reflection direction of the ion beam and the standing angles are nearly same as the ion incidence angle with the surface normal. This nano-pillas are easily transferred from the InP surface to double sided carbon tape without any damage. High Resolution Transmission Electron Microscopy (HRTEM) study of single nano-pillar reveals that this nano-pillar are almost crystalline in nature except 2-4 nm amorphous layer on the outer surface. The transmission electron microscopy combined with energy-dispersive x-ray spectroscopy (TEM-EDS) analysis of these nano pillars exhibit that the ratio of In and P is little higher compared to the bulk InP.

Email: dipakiop@gmail.com

# Introduction:

Fabrication of regular arrays of semiconductor nano structures is of immense importance due to their potential applications in optoelectronic, photonic and recording media [1]. Compared to gallium arsenide (GaAs), indium phosphide (InP) has number advantageous properties such as higher breakdown voltage, higher electron mobility, higher thermal conductivity, smaller ionization coefficients and dielectric constants[2]. Fabrication of InP nano-pillars through self-organization induced by ion irradiation processes is of special interest due to the possibility of production of regular arrays of pillars on large areas in a single technological step[3]. Until now several works have been reported on dry etching processes aiming at such fine structures [4-7] . It has been shown that ECR ion beam sputtering can be a promising tool for nano scale pattering on semiconductor surfaces [8-9]. This sputtering technique causes very low damage or amorphization to the target material. The nanostructures formed by this sputtering method can be almost crystalline except a few nm of the surface wall are amorphous.

Recently, there are several report of formation of regular nano structures on single crystalline semiconductor surfaces due to low energy $Ar^+$ ion sputtering processes [10-14] The reason for such kind of pattern formation are described as due to the competition of curvature dependant ion sputtering that roughens the surface and smoothening by different relaxation mechanisms. However, most of these studies report the formation of nano dots or ripples on the sputtered surfaces.

Here we report the fabrication of highly dense, self-assembled and uniform nano pillars and cones by ECR $Ar^+$ ion sputtering on InP(111) surfaces. The structures were fabricated without using any mask or templates. From the analysis of the SEM images it is found that these nano pillars having height about 500 nm and width about 80 nm and therefore, the aspect ratio of these nano pillars are greater than 6. The standing-angle of these nano pillars measured with respect to the surface normal is almost same as the ion incidence angle with respect to the surface normal and the growth direction of the nano pillars is along the reflection of ion beam from the surface. Detailed Transmission Electron Microscopy (TEM) analysis reveals that these nano pillars are having the same crystalline structure as InP while the outer surface wall of width 2-3 nm are amorphous in nature.

## Experimental

Undoped and mirror polished InP(111) single crystal wafer were irradiated with ECR $Ar^+$ ion inside a chamber with base pressure $10^{-7}$ Torr. The irradiation of $Ar^+$ ion was done at 1 keV energy and with ion flux of $10^{16}$ ions/ $cm^2$ / sec. The irradiation was done at various ion incidence angles 15º, 30º, 45º, 60º , 75º with respect to the sample surface normal. Irradiation was done for three different time durations of 100, 500 and 1000 sec at each incidence angles mentioned above. The sample temperature during irradiation was measured and it was found that sample temperature is higher for near normal incidence compare to near grazing incidence. The substrate temperature varies between 150º C to 200º C. The ion irradiated surface shows the growth of self assembled InP nano pillars. The nano pillars were investigated in details by utilizing high resolution scanning electron microscope (SEM). The crystal structure and chemical composition of the nano pillars were investigated by utilizing high resolution transmission electron microscopy combined with energy-dispersive x-ray spectroscopy (TEM-EDS) technique.

## Results and Discussion

Figure1 shows the SEM images of the InP(111) surfaces irradiated with 1 keV $Ar^+$ ion from ECR source for different incident angles 15º, 60º, 75º and 85º for 1000 secs, with ion flux of $10^{16}$ ions/ $cm^2$ / sec. Fig1a shows the evolution of the topography of InP(111) surface due to the Ar ion irradiation at 15˚ incident angle with respect to the surface normal. The figure shows the appearance of irregular random structures and there is no clear formation of nano–pillars on the surface. Figure 2b shows the surface topography when the ion incident angle is 60 degree with respect to the normal of the InP(111) surface. In this sputtering condition it is seen that a regular pattern of nano pillar structure are formed on the InP surface. These nano pillars having height about 500 nm and width about 80 nm and so the aspect ration of these pillars are larger than 6. Interestingly, these nano pillars also make an angle of 60˚ with the surface normal. Subsequently it is observed that this angle of the nano pillar with the surface normal (A) changes with the ion beam inceident angle (B). The value of the first angle (A) is almost same with the second angle(B). When the incident angle of the ion irradiation is changed to 75 degree, the fabricated nano pillar grow in nearly same angle with the InP surface normal. Nano-pillars are also be fabricated which are lying flat on the surface (fig. 2d) of InP(111) by irradiating the surface with 85 degree angle w.r.t the surface normal. It seems the growth direction of the nano pillars is same as the reflection direction of the incident ion from the surface.

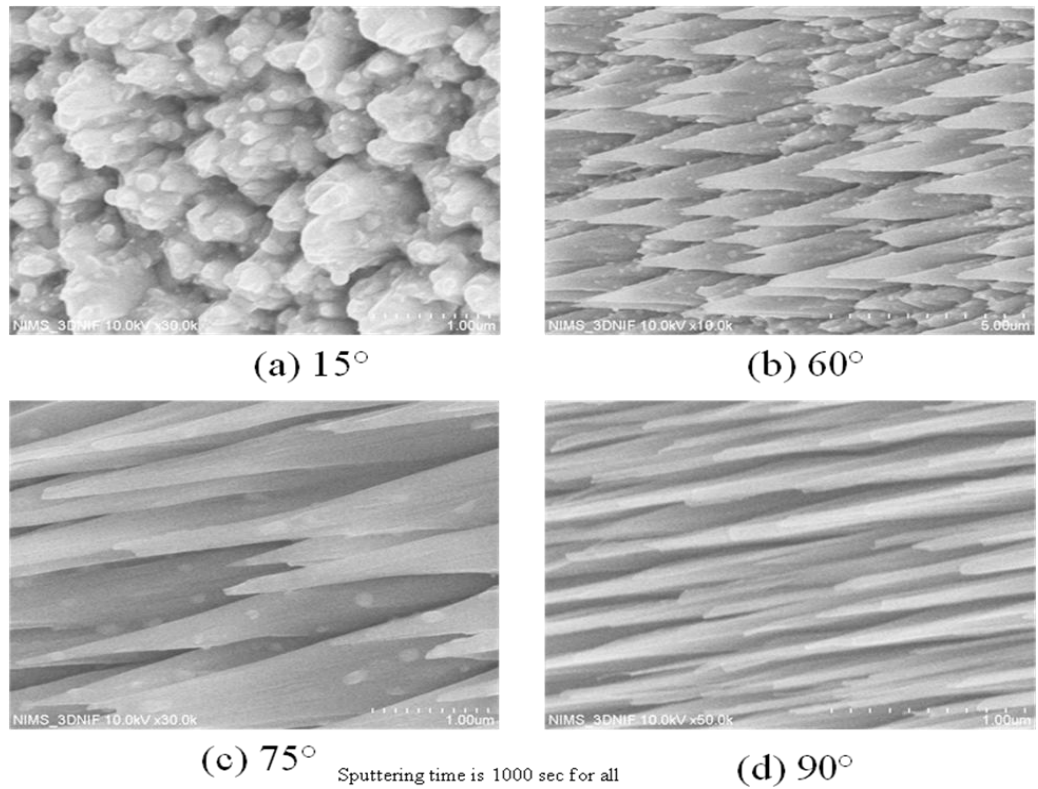

Figure 1: Shows the change in orientation of the nano-pillars with the incident angle (a) 15, (b) 60, (c) 75 and (d) 90 degrees with respect to the surface normal.

The evolution of the nano pillars for varying ion irradiation time (fluence) while keeping the ion incident angle fixed at 60° are shown shown in figure 2. Figure2a show the SEM image InP(111) surface irradiated with $Ar^+$ ion for 100 sec which corresponds to the fluence of $1\times10^{18}$ ions/cm$^2$ . The figure shows formation of very regular, self organized nano pillars with sharp tips. The average height of these nano pillars is about 600 nm and width 80 nm. So, these nano pillars are of high aspect ratio which is greater than 7. However, the width of the nano pillar is not uniform over the whole length, the bottom is wider and the top is sharper, like a extended cone. As it is clearly visible from the figure 2b, there is a tiny In metallic Indium ball, which gradually becomes smaller and disappears for longer sputtering for 500 sec as shown in Fig 2b. Nozu et al. [15] and Lee at el. [16] have reported the formation of cone formation with Indium metallic ball on top of the cone due to ion sputtering. But their study does not show very regular and densely packed nano pillars as observed here. Fig 2b shows the nano-pillar formation for 500 sec of irradiation (fluence $5\times10^{18}$ ions/cm$^2$ ). It shows the most regular in shape and densely packed nano-pillars with uniform height and width. The height of these nano pillars are about 500 nm and width 50 nm. The shape of the pillars are like a rods having uniform width from top to bottom. However after 1000 sec of irradiation as shown in fig 2c, nano pillar with sharp tip and cone type of structure are observed again as in the case of 100 sec. The height of these nano pillars are about 450 nm and average width is 60 nm and so the aspect ration is more than 7. It is observed that the height of the nano pillar decreases with increasing sputtering time.

Interestingly, all these nano pillars can be easily removed from the InP(111) surface by cleaving with adhesive carbon tape. After cleaving the nano-pillar with double sided carbon tape, it is found that the nano pillar can stand on the carbon tape with the top side down. Figure 2d shows SEM images of the top side down vertically standing nano pillars on the carbon tape.

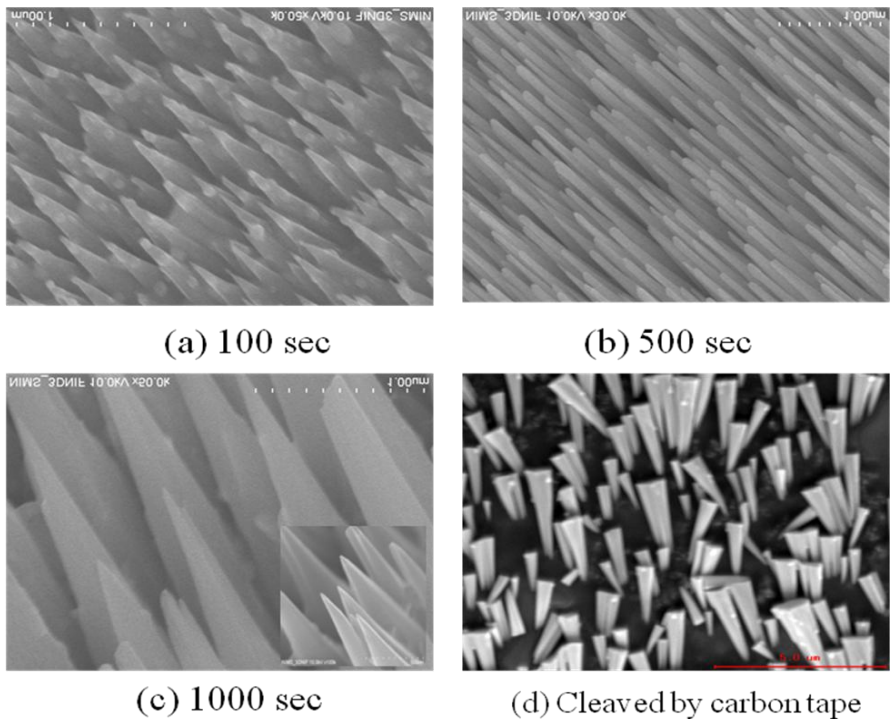

Figure 2: SEM images shows the shape of the nano- pillars with the irradiation time (a) 100 sec, (b) 500 sec and (c) 1000 sec (inset shows the high resolution view of sharp tips of the pillars). Fig 3(d) shows the cleaved structures reverse-vertically standing on carbon tape.

Figure 3 shows the HRTEM images of these InP nano-pillars. Fig 3a shows the side view of two InP nano pillars on a Cu grid. The high resolution TEM (HRTEM) image of one of this nano-pillars is shown in fig 4b. The HRTEM image shows that except 2-3 nm outside amorphous layer the most part of the nano pillar are crystalline in nature. The crystalline nature are confirmed by transmission electron diffraction (TED) pattern shown in the inset of fig. 4(b). The TED pattern shows the <111> growth direction of the pillar with the zinc blend structure of InP.

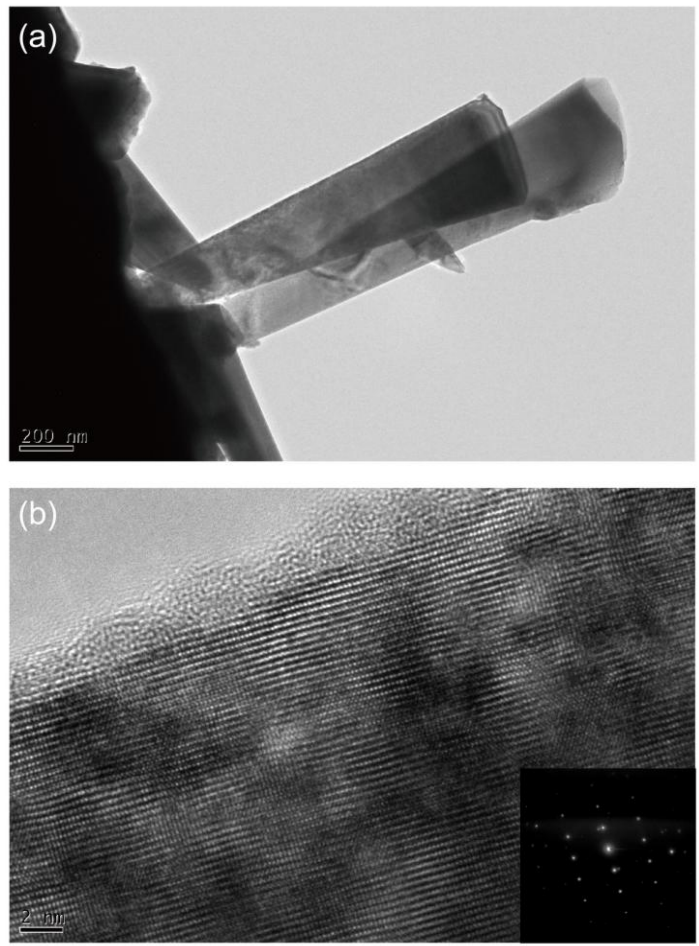

Figure 3: TEM image of InP nanorods on a Cu grid (a). A lattice image and the corresponding electron diffraction pattern are displayed in Fig. (b). The thickness of the outer surface amorphous layer is about 2 - 3 nm.

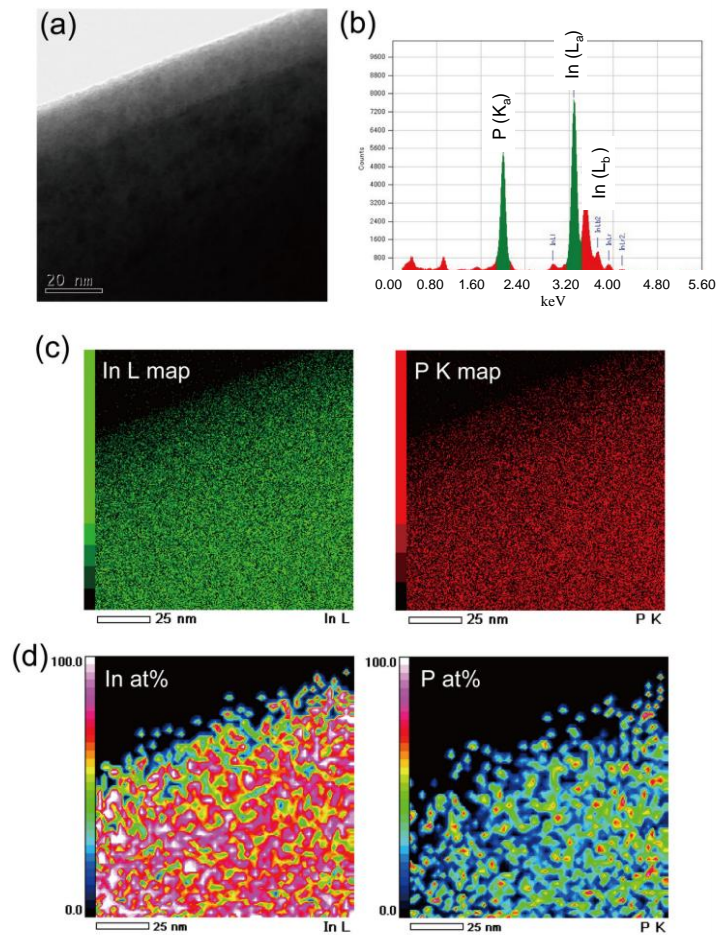

**Figure 4: EDS mapping was performed for an area shown in Fig. 4(a) which is the same area as in fig 3(b) (STEM-bright field image). Figure 4(b) shows an energy-dispersive spectrum of X-ray generated by the entire area, giving the atomic ratio of In: 62.25% and P: 37.75%. An surface In (metallic) layer can not be detected as seen in the In (L) and P (K) maps (Fig. 4(c)). The quantitative maps indicate a short-range fluctuation in the atomic ratio (Fig. 4(d))**

Two-dimensional X-ray EDS elemental mapping was recorded from the same area as in fig. 3a and are shown in fig 4a. This is a bright-field scanning transmission electron microscopy (STEM) image. The P K-line and In L-lines used in the mapping are shown in Fig 4b. The X-ray imaging of In L-line and P K-lines are shown in Fig. 4c. The close analysis of these images reveals that the InP nano pillars having In concentration 62.25 % and P concentration of 37.75 %. The formation of this InP nano pillar can be understood on the basis of preferred sputtering of P atoms leading to the enriched In atoms on the surface. This acces In atoms agglomerate to form In islands. The sputtering yeild of In is less than that of InP. Thus the InP surface is sputtered faster than the regions of metallic In, leading to the formation of sputter cones or pillars[15,16].

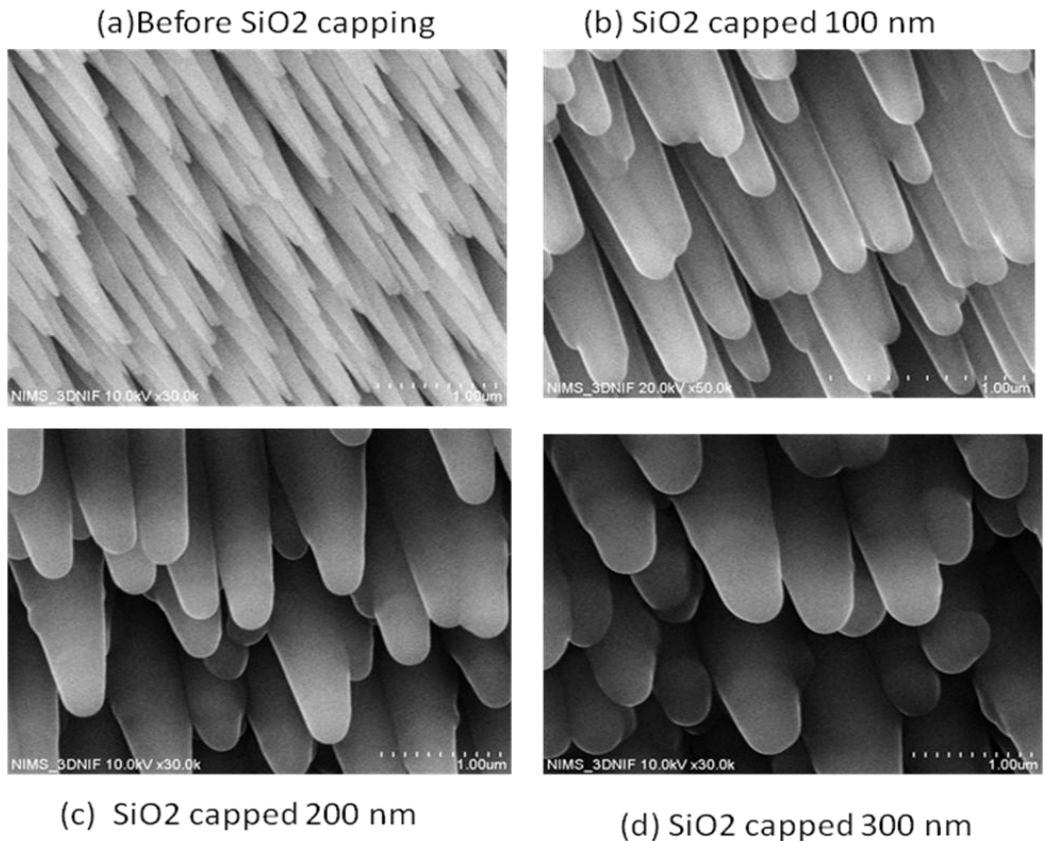

Figure 5: (a) Shows the InP nano-pillar structures after 1 keV Ar ion irradiation for 1000 sec in . This nano pillar are used as template for the growth of SiO2 hollow nano pillar of different wall thickness (b) 100 nm, (c) 200 nm and (d) 300 nm

One of the applications of these fabricated nano pillars shown in fig. 5a may be to use it as template to grow other materials. Here we show in fig 5 a,b,c that $SiO_2$ nano pillars of different thickness has been grown by PECVD method using these InP nano pillars as core structures. These kind of $SiO_2$ nano structures can have immense technological applications for solar cells and moth eye type of antireflective structures[17,18].

## Summary & Conclusion:

We have fabricated self organized InP nano pillars using electron cyclotron resonance Ar ion irradiation. The growth of the nano pillars are along the reflection direction of the incident ion beam. The bulk of nano pillars are crystalline in nature except 2-3 nm of the outer surface is amorphous. These nano pillars easily transferable to other surfaces and can be used as template to grow other materials for technological applications.

## Acknowledgement:

One of the author (D.P.) would like to thank Dr Masaki Ogawa for useful discussions and for help in planning experiment. This work was partially supported by SENTAN-JST.

# References:

1. Andrew N. Shipway, Eugenii Katz, Itamar Willner, Chem Phys Chem, **1**, 18 (2000).
2. D. Streit, Compound Semiconductors, May 2002; B. Humphreys and A. O'Donell, Compound Semiconductors, August 2003; D. Lammers, Electronic Engineering Times, 12 September 2002.
3. D. Paramanik, S.N. Sahu and S. Varma, J. Phys. D: Appl. Phys. **41**, 125308 (2008).
4. B. E. Maile, A. Forchel, R. Germann and D. Grutzmacher: Appl. Phys. Lett. **54** (1989) 1552.
5. K. Kudo, Y Nagashima, M. Tamura, S. Tainura, A. Ubukata and S. Arai: Jpn. J. Appl. Phys. **33** (1994) L1382.
6. M. Tamura, Y Nagashima, K. Kudo, K. C. Shin, S. Tamura: A. Ubukata and S. Arai: Jpn. J. Appl. Phys. **34** (1995) 3307.
7. U. A. Griesinger, H. Schweizer, S. Kronmuller. M. Geiger, D. Ottenwalder. F Scholz and M H. Pilkuhn: IEEE Photon. Technol. Lett. **8** (1996) 583.
8. B. Jacobs, M. Emmerling. A. Forchel, I. Gyuro. P. Speier and E. Zielinski: Jpn. J. Appl. Phys. **32** (1993) L173.
9. G. A. Vawter and C. I. H. Ashby: J. Vac. Sci. Technol. B **12** (1994) 3374.
10. Peter Sigmund, Phys. Rev. **184**, 383 (1969).
11. R. Mark Bradley and James M. E. Harper, J. Vac. Sci. Technol. A **6**, 2390 (1988).
12. C. M. Demanet et al., Surf. Interface Anal. **23**, 433 (1995); **24**, 497 (1996), **24**, 503 (1996).
13. B. Kahng, H. Jeong and A.-L. Barabasi, Appl. Phys. Lett. **78**, 805 (2001).
14. S. Facsko, T. Dekorsy, C. Koerdt, C. Trappe, H. Kurz, A. Vogt, H. L. Hartnagel, Science **285**, 1551 (1999).
15. M. Nozu, M. Tanemura, M Okayauma, Surf. Sci. Lett. **304**, L468 (1994).
16. Hyung-Ik Lee, Tomoki Akita, Ryuichi Shimizu, Surf. Sci **412**, 24 (1998).
17. Stuart A. Boden and Darren M. Bagnall, Prog. Photovolt: Res. Appl. **18**, 195 (2010)
18. Huang YF, Chattopadhyay S, Jen Y-J, Peng C-Y, Liu T-A, Hsu Y-K, Pan C-L, Lo H-C, Hsu C-H, Chang Y-H, Lee C-S, Chen K-H, Chen L-C., Nature Nanotechnology **2**, 770 (2007)